\documentclass{PoS}
\usepackage{latexsym,graphicx,epsfig,cite}
\usepackage{amsmath}
\usepackage{amssymb}
\newcommand{\half}{{\textstyle\frac{1}{2}}}
\newcommand{\hc}{\hbox {h.c.}}

\renewcommand{\Im}{{\rm Im\thinspace}}
\def\gsim{\mathrel{\rlap{\raise 2.5pt \hbox{$>$}}\lower 2.5pt\hbox{$\sim$}}}

\title{Update on the CP-Violating Inert-Doublet Model}

\ShortTitle{The CP-Violating Inert-Doublet Model}

\author{B. Grzadkowski\\
        Institute of Theoretical Physics, Faculty of Physics, University of Warsaw, \\ Ho\.za 69, 
PL-00-681 Warsaw, Poland \\
        E-mail: \email{Bohdan.Grzadkowski@fuw.edu.pl}}

\author{O. M. Ogreid\\
        Bergen University College, Bergen, Norway \\
        E-mail: \email{omo@hib.no}}

\author{\speaker{P. Osland}%
 \\
       Department of Physics,
University of Bergen, Postboks 7803, N-5020 Bergen, Norway\\
       E-mail: \email{per.osland@ift.uib.no}}

\author{A. Pukhov\\
        Skobeltsyn Inst.\ of Nuclear Physics, Moscow State Univ., Moscow 119991, Russia \\
        E-mail: \email{pukhov@lapp.in2p3.fr}}

\author{M. Purmohammadi\\
        Department of Physics,
University of Bergen, Postboks 7803, N-5020 Bergen, Norway \\
        E-mail: \email{Mahdi.PurMohammadi@ift.uib.no}}
        
\abstract{We have updated a recently proposed extension of the Inert Doublet Model. The extension amounts to the addition of an extra non-inert scalar doublet. The model thus offers a possibility of CP violation in the scalar sector and a candidate for the Dark Matter. The recent XENON100 direct-detection experiment excludes a considerable range of medium--low dark-matter masses, leaving only as viable very low masses of order 5--10~GeV, as well as the regions from $\sim 60$ to $\sim110$~GeV, and above $\sim530$~GeV. For favorable parameter regions one may observe related long-lived charged particles produced at the LHC.}

\FullConference{The XXth International Workshop  High Energy Physics and Quantum Field Theory  \\
		  September 24 - October 1,  2011 \\
		 Sochi, Russia}

\begin{document}

\section{The model}
The Inert Doublet Model (IDM) \cite{Deshpande:1977rw,Barbieri:2006dq} provides a very economical extension of the Standard Model, allowing for Dark Matter. It is basically a Two-Higgs-Doublet Model \cite{HHG}, where one doublet is protected from having a vacuum expectation model by virtue of an imposed $Z_2$ symmetry. The resulting spectrum of scalars has an ``ordinary'' neutral Higgs particle, $H$, a pair of charged ones, and two additional neutral ones, which we shall denote $S$ and $A$. The lighter of these, assumed to be the scalar $S$, is then the dark matter. In addition to providing an economical accommodation of dark matter, the model also offers a mechanism for generating neutrino masses \cite{Deshpande:1977rw} and alleviates the ``Little hierarchy'' \cite{Barbieri:2006dq} by allowing Higgs masses as high as 400~GeV.

The IDM has been studied extensively \cite{Cirelli:2005uq,LopezHonorez:2006gr,Hambye:2007vf,Cao:2007rm,Andreas:2008xy,Lundstrom:2008ai,Hambye:2009pw,LopezHonorez:2010tb}, and two mass regions have been identified, a low-mass region, from about 5~GeV to about 110~GeV, and a high-mass region, beyond about 535~GeV. Above some 110~GeV, the annihilation in the early Universe, to two gauge bosons ($W^+W^-$ or $ZZ$) becomes very fast, and the DM density would be too low. Eventually, for sufficiently heavy DM particles (above approximately 535~GeV), the annihilation rate drops sufficiently for the remaining DM density to again become compatible with the data.

In the extension discussed here, the CP-violating Inert-Doublet Model \cite{Grzadkowski:2009bt,Grzadkowski:2010au}, an additional doublet is added, for a total of two non-inert doublet plus an inert one with no vacuum expectation value, and hence no Yukawa couplings. The additional non-inert doublet allows for CP violation in the scalar sector, which is desirable for cosmological reasons. On the other hand, this additional feature has a price: there are more parameters than in the IDM.

Denoting the non-inert doublets $\Phi_1$ and $\Phi_2$, and the inert one $\eta$, we take the scalar couplings to be given by
\begin{align}
V_{123}(\Phi_1,\Phi_2,\eta)
&=
\lambda_{a} (\Phi_1^\dagger\Phi_1)(\eta^\dagger \eta)
+\lambda_{a} (\Phi_2^\dagger\Phi_2)(\eta^\dagger \eta) \nonumber  \\
& +\lambda_{b}(\Phi_1^\dagger\eta)(\eta^\dagger\Phi_1)
+\lambda_{b}(\Phi_2^\dagger\eta)(\eta^\dagger\Phi_2) \nonumber  \\
&
+\half\left[\lambda_{c}(\Phi_1^\dagger\eta)^2 +\hc \right]
+\half\left[\lambda_{c}(\Phi_2^\dagger\eta)^2 +\hc \right].
\label{Eq:v123}
\end{align}
Here, ``dark democracy'' has been imposed, the two non-inert doublets couple in the same way to the inert doublet. Also, $(\Phi_1,\Phi_2)$ and $\eta$ are subject to standard quartic potentials.

The resulting scalar spectrum is as follows: In the 2HDM sector we have three neutral Higgs bosons ($H_1$, $H_2$, $H_3$), and two charged ones ($H^\pm$), whereas in the inert sector we have two neutral ones ($S$ and $A$) and a pair of charged ones ($\eta^\pm$). We assume $S$ to be the lightest of these, and thus the dark matter. While they have no couplings to fermions, they do couple to gauge bosons, and to the Higgs sector via the potential (\ref{Eq:v123}).
We specify the mass spectrum, rather than the potential, and find
\begin{subequations} 
\label{Eq:lambda-vs-splitting}
\begin{align}
\lambda_a&=\frac{2}{v^2}
\left(M^2_{\eta^\pm}-m_\eta^2\right), \\
\lambda_b&=\frac{1}{v^2}\left(M^2_S+M^2_A-2M^2_{\eta^\pm}\right), \\
\lambda_c&=\frac{1}{v^2}\left(M^2_S-M^2_A\right),
\end{align}
\end{subequations}
where $m_\eta$ is a mass parameter of the $\eta$ potential  \cite{Grzadkowski:2009bt}.

The coupling of the inert particles to the Higgs sector is largely controlled by
\begin{equation} \label{Eq:lambda_L}
\lambda_L\equiv \half(\lambda_a+\lambda_b+\lambda_c)
=\frac{M_S^2-m_\eta^2}{v^2},
\end{equation}

\section{Phenomenology}
At low $S$ mass, the annihilation in the early Universe proceeds via the coupling to the lightest neutral Higgs boson to $c\bar c$ and $b\bar{b}$. This can easily be made compatible with the DM density, via a tuning of the Higgs mass and the couplings involved. At higher $S$ mass, the $W^+W^-$ threshold opens up. The coupling to the gauge bosons can not be tuned, and at some point (for some critical value of $M_S$) the annihilation becomes too fast to agree with the DM density.
Theoretical and experimental constraints were imposed. The former include positivity, unitarity, and electroweak symmetry breaking, whereas the latter include $b\to s\gamma$, $B\bar B$ oscillations, $\Gamma(Z\to b\bar b)$, electroweak precision data and the electron electric dipole moment \cite{Grzadkowski:2010au}. As a lowest bound on the $\eta^\pm$ mass, we take the LEP chargino bound, $M_{\eta^\pm}>70~\text{GeV}$ \cite{Pierce:2007ut}. Also $M_A$ is constrained by LEP data, to $M_A\gsim110~\text{GeV}$ \cite{Lundstrom:2008ai}. A detailed scan over the many parameters was performed, and allowed regions were identified \cite{Grzadkowski:2010au}. 

A dedicated implementation of the model was made in the {\tt micrOMEGAs} software \cite{Belanger:2006is,Belanger:2008sj}. This allows to determine the Early-Universe DM density \cite{Hinshaw:2008kr}, which plays an important role in constraining the allowed parameter space. The same software also allows determining the direct-detection cross section, shown in Fig.~\ref{Fig:exclusion} together with exclusion limits from CDMS-II \cite{Ahmed:2009zw} and XENON100 \cite{Aprile:2011hi}.

\begin{figure}[h]
\refstepcounter{figure}
\label{Fig:exclusion}
\addtocounter{figure}{-1}
\begin{center}
\setlength{\unitlength}{1cm}
\begin{picture}(14.0,7)
\put(3.0,0.)
{\mbox{\epsfysize=7.5cm\epsffile{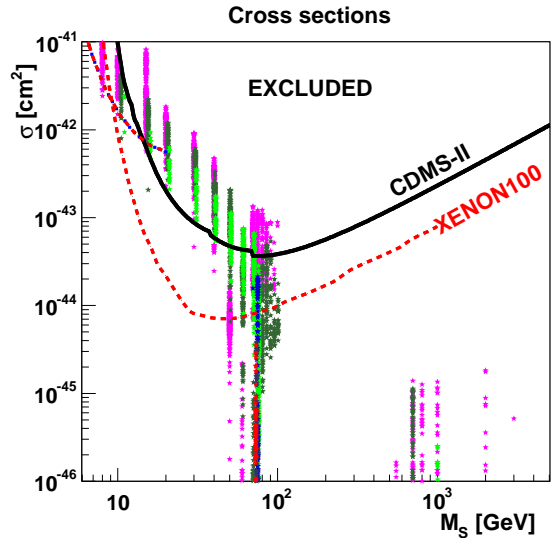}}}
\end{picture}
\caption{Direct-detection cross sections for selected DM masses $M_S$. Different colors refer to different Higgs masses. Magenta: $M_1\leq120~\text{GeV}$, green: $150~\text{GeV}\leq M_1\leq230~\text{GeV}$, blue: $300~\text{GeV}\leq M_1\leq400~\text{GeV}$, red: $M_1\geq500~\text{GeV}$. The region from $M_S\sim110~\text{GeV}$ to $\sim550~\text{GeV}$ is not shown, as the model there yields too low DM-density.}
\end{center}
\end{figure}

While the 2010 exclusion limits \cite{Ahmed:2009zw,Aprile:2010um} were marginally compatible with the model over essentially the whole range of $M_S$ values, this is no longer true for the most recent XENON100 data \cite{Aprile:2011hi}, which exclude the range of $M_S$ values from around 10~GeV to around 60~GeV. Since the present model has more parameters than the simple IDM, this exclusion applies also to that model.

Anyway, the most interesting region is for $M_S\sim75~\text{GeV}$, where the Little hierarchy can be significantly alleviated, by allowing $M_1$-values up to 400--600~GeV \cite{Barbieri:2006dq,Grzadkowski:2010au}.

\section{LHC prospects}
The gauge coupling of the inert doublet permits pair production in $pp$ collisions at the LHC,
\begin{equation}
pp\to SSX, AAX, SAX, S\eta^\pm X, A\eta^\pm X, \eta^+\eta^- X.
\end{equation}
The $A$ and $\eta^\pm$ would subsequently decay to the lightest one, $S$. 

The decay (via a virtual $W$)
\begin{equation} \label{Eq:eta-decay}
\eta^+ \to S \ell^+\nu_\ell
\end{equation}
is similar to the muon decay, except that (i) a scalar-scalar-vector vertex replaces a fermion-fermion-vector vertex, and (ii) one of the invisible final-state particles is massive. For $M_{\eta^\pm}-M_S\ll M_{\eta^\pm}$, the decay rate can be written as
\begin{equation} \label{Eq:Gamma_eta}
\Gamma_{\eta^\pm}=\frac{G_\text{F}^2}{30\pi^3}\left(M_{\eta^\pm}-M_S\right)^5.
\end{equation}
If the mass difference is small, this rate could be suppressed, and the $\eta^\pm$ give a visible track in the detector. In the high-mass region, $M_S\gsim550~\text{GeV}$, these masses are necessarily rather degenerate, because of positivity and the DM constraint.

The region of more immediate interest is the one around $M_S=75~\text{GeV}$, since such particles would have a significant cross section for being pair produced at the LHC. For $M_1=120~\text{GeV}$ and $M_S=75~\text{GeV}$, we show in Fig.~\ref{Fig:cross-section} the inclusive $\eta$ pair production cross section.

\begin{figure}[h]
\refstepcounter{figure}
\label{Fig:cross-section}
\addtocounter{figure}{-1}
\begin{center}
\setlength{\unitlength}{1cm}
\begin{picture}(14.0,7)
\put(3.0,0.)
{\mbox{\epsfysize=7cm\epsffile{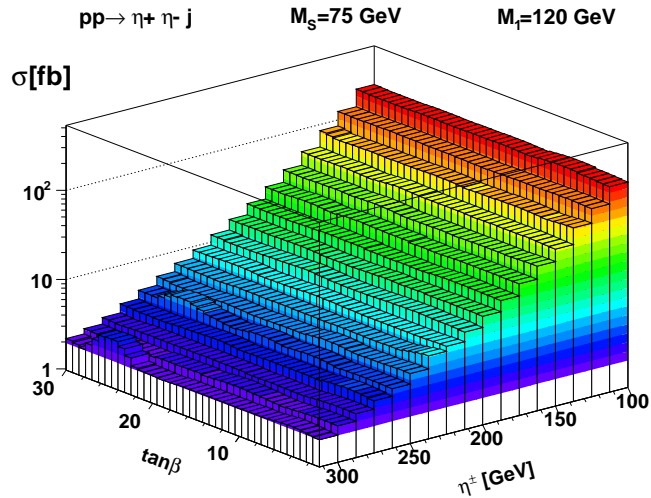}}}
\end{picture}
\caption{$pp\to(\eta^+\eta^-\text{jet})$ cross section at $\sqrt{s}=7~\text{TeV}$.}
\end{center}
\end{figure}

Alternatively, if $A$ particles are produced, these could decay as
\begin{equation}
A\to SZ^*\to Sf\bar f,
\end{equation}
leading to lepton pairs or quark jets that might yield observable signals.

Such an inert sector would also modify the Higgs branching ratios, if decay to these scalars are kinematically possible. This is illustrateted in Fig.~\ref{Fig:higgs-branchings} for the case of
\begin{equation} \label{Eq:dark-masses}
M_S=75~\text{GeV}, \quad M_{\eta^\pm}=85~\text{GeV},\quad M_A=110~\text{GeV}.
\end{equation}
A crucial parameter here is the $H_jSS$ trilinear coupling, $\lambda_L$, for which two values are considered in the figure, both compatible with the theoretical and experimental constraints.

\begin{figure}[h]
\refstepcounter{figure}
\label{Fig:higgs-branchings}
\addtocounter{figure}{-1}
\begin{center}
\setlength{\unitlength}{1cm}
\begin{picture}(14.,6.5)
\put(-1,0.)
{\mbox{\epsfysize=6.2cm\epsffile{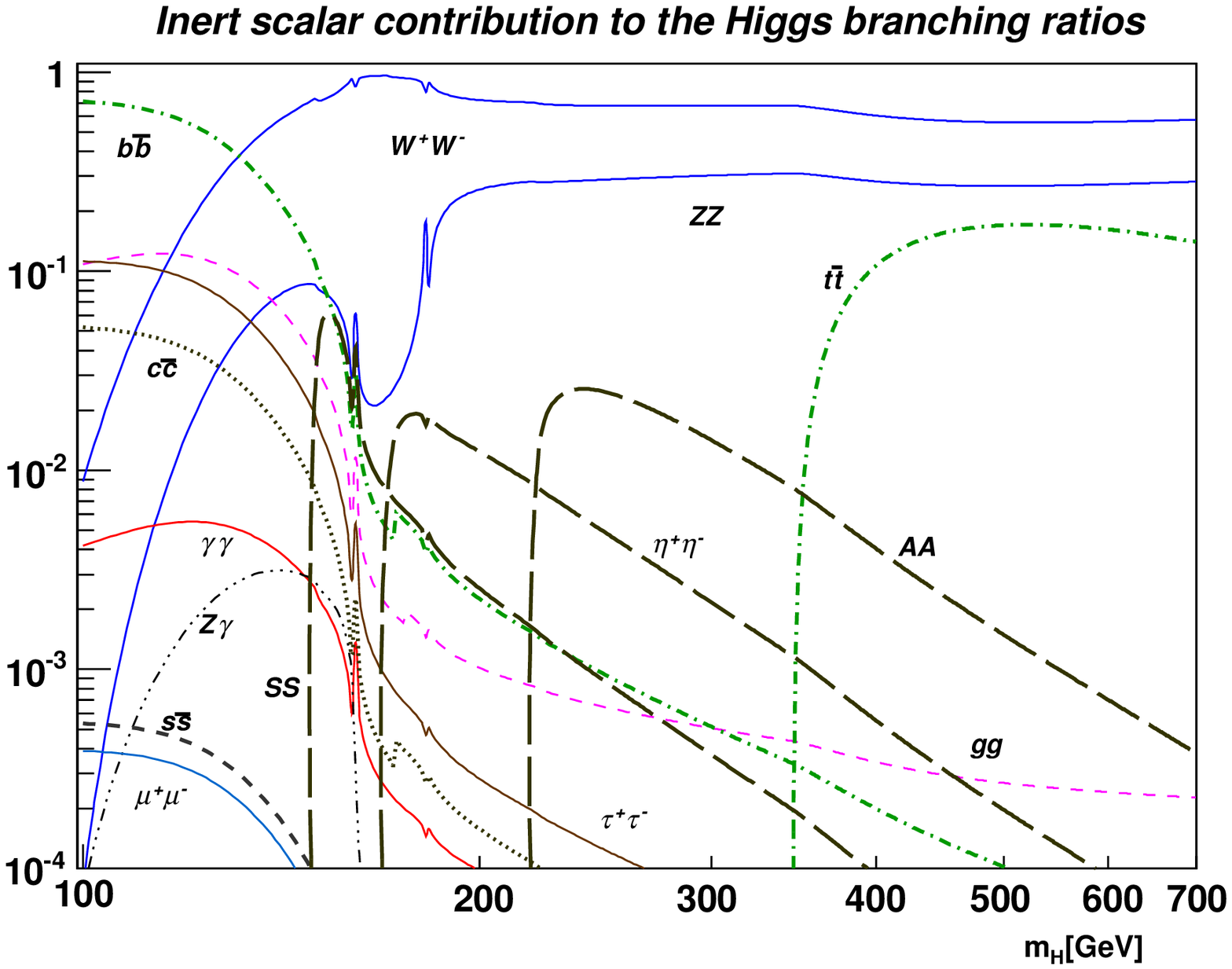}}
\hspace*{-8mm}
 \mbox{\epsfysize=6.2cm\epsffile{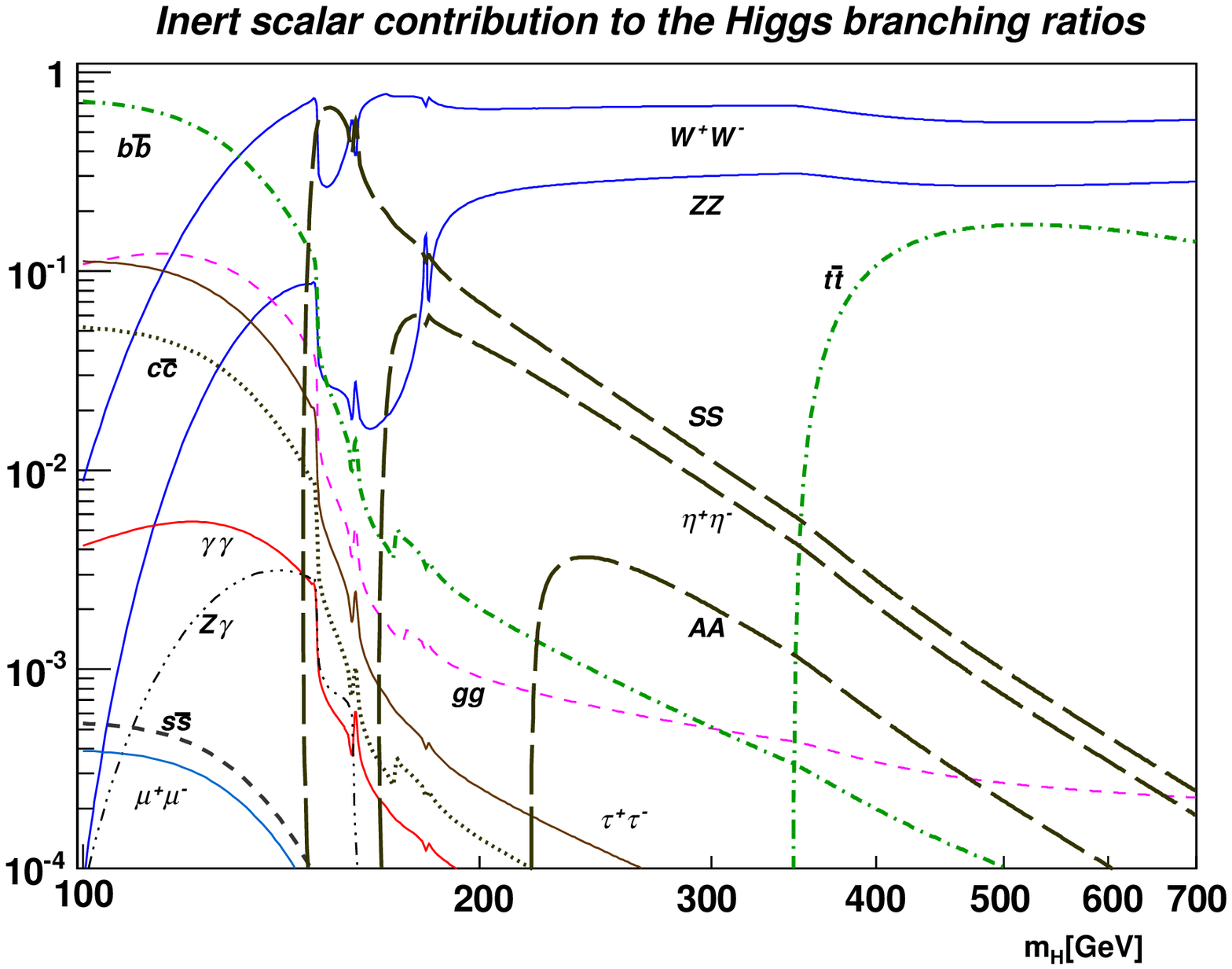}}}
\end{picture}
\caption{Modified Higgs branching ratios, for the masses given above and two values of $\lambda_L$; left: $\lambda_L=0.018$, right: $\lambda_L=0.21$.}
\end{center}
\end{figure}
For the case of current interest, $M_1\simeq125~\text{GeV}$, the marginally allowed DM mass of $M_S\simeq60~\text{GeV}$ could significantly reduce the $H_1$ branching ratios to SM particles.

\section{CP violation}
To illustrate the amount of CP violation that is available in the model, we 
consider imaginary parts of weak-basis-transformation invariants that are sensitive to CP violation in the scalar potential. The
advantage of studying invariants stems from the fact that they offer a realistic measure of CP violation
since any CP-violating observable that emerges from the scalar potential must be a linear combination of the invariants (or their higher odd powers).
In a 2HDM there are three
independent invariants $J_{1,2,3}$ that are sufficient to describe any CP-violating phenomenon. For three doublets one should expect more invariants, however here for illustration, we limit ourselves to only the $J_{1,2,3}$ 
defined in \cite{Gunion:2005ja} for a 2HDM. 

\begin{figure}[h]
\refstepcounter{figure}
\label{Fig:cp-violation}
\addtocounter{figure}{-1}
\begin{center}
\setlength{\unitlength}{1cm}
\begin{picture}(14.0,7)
\put(1.0,0.)
{\mbox{\epsfysize=7cm\epsffile{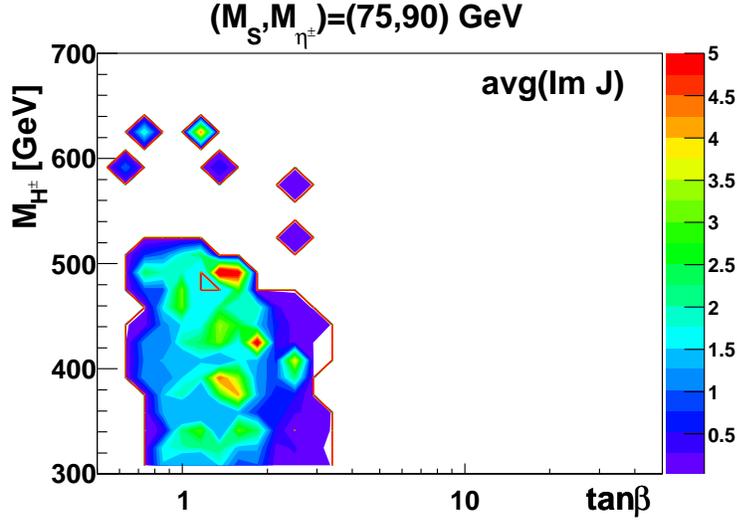}}}
\end{picture}
\caption{Contour plots for averaged (over $\alpha$'s) imaginary parts of the algebraic average of the invariants, $\Im J_{1,2,3}$, illustrating the strength of CP violation, vs $\tan\beta$ and $M_H^\pm$, for $(M_S,M_A,M_{\eta^\pm},m_\eta)=(75,110,90,100)~\text{GeV}$, and $(M_1,M_2,\mu)=(120,300,200)~\text{GeV}$.}
\label{Im-J_i-075-090}
\end{center}
\end{figure} 

It turns out that $\Im J_{1,2,3} \sim 0.5 - 3$,
that is five orders of magnitude more than the corresponding (Jarlskog) invariant  
in the SM. We show in Fig.~\ref{Fig:cp-violation} the average of these quantities,
\begin{equation}
\text{avg}(\Im J)
=\frac{1}{3}(\Im J_1 + \Im J_2 + \Im J_3),
\end{equation}
for the choice of parameters $M_S=75~\text{GeV}$ and $M_1=120~\text{GeV}$ (for details, see \cite{Grzadkowski:2010au}). The soft mass parameter $\mu$ is taken to be 200~GeV. The allowed region is  restricted to $\tan\beta={\cal O}(1)$, larger values can be reached by tuning $\mu$.

\section{Summary}

The Inert Doublet Model has recently received a lot of attention, in large measure because of its simplicity. We have shown that the extension to an extra doublet, which permits some amount of CP violation, is in conflict with the XENON-100 data for a wide range of DM masses, from about 10 to 60~GeV. Consequently, also the IDM is excluded in this region, as recently also pointed out  \cite{Djouadi:2011aa} in a more general context. However, at higher masses, the model is viable. In particular, the region around $M_S=75~\text{GeV}$ remains very interesting.

\end{document}